\documentclass[12pt,a4paper,twoside]{article}
\usepackage{epsfig}
\usepackage{graphics}
\newlength{\figwidth}
\newlength{\figheight}
\def\pyver{6.214}

\def\gaga{ \gamma \gamma }
\def\bbbar{ b\overline{b} }
\def\ccbar{ c\overline{c} }

\def\hbb{ h \rightarrow \bbbar }

\def\gagahbb{ \gaga \rightarrow \hbb }

\def\gagabb{ \gaga \rightarrow \bbbar }
\def\gagacc{ \gaga \rightarrow \ccbar }
\def\gagabbcc{ \gagabb , \, \ccbar }

\def\gagabbg{ \gagabb (g) }
\def\gagaccg{ \gagacc (g) }
\def\gagabbgccg{ \gagabbg , \, \ccbar(g) }

\def\sgagahbb{ \sigma( \gagahbb ) }

\def\Ghgagahbb{ \Gamma (h \rightarrow \gaga ){\rm Br}( \hbb ) }

\def\Wgaga{ W_{\gaga} }

\def\epem{ e^{+} e^{-} }
\def\emem{ e^{-} e^{-} }

\def\Mh{ M_h }
\def\Mheq{$ \Mh = $ }

\def\sqrtsee{ \sqrt{s_{ee}} }
\def\sqrtseeeq{$ \sqrtsee  = $ }

\newcommand{\gagahad}{\( \gaga \rightarrow \) {\em hadrons}}

\newcommand{\bbtagging}{\( \bbbar  \)-tagging}
\newcommand{\btagging}{\( b \)-tagging}

\newcommand{\ccmistagging}{\( \ccbar  \)-mistagging}

\def\z0{Z}
\def\eg{{e.g.} }
\def\ie{{i.e.} }

\title{
       \vspace{-1.5cm}
       \begin{flushright}
       \begin{tabular}{l}
       {\large IFT - 19/2003}    \\[-3mm]
       {\large hep-ph/0307183 }\\[3mm]
       {\large July 2003}
       \end{tabular}
       \end{flushright}
\vspace{1.5cm}
Light~Higgs-boson~production
at~the~Photon~Collider~at~TESLA \\
with an improved background analysis
}
 \author{ \\[1cm]
  Piotr Nie\.zurawski, Aleksander Filip \.Zarnecki \\
 {\small\it Institute of Experimental Physics, Warsaw University, 
    ul. Ho\.za 69, 00-681 Warsaw, Poland} \\[3mm]
 Maria Krawczyk \\
 {\small\it Institute of Theoretical Physics, Warsaw University, 
        ul. Ho\.za 69, 00-681 Warsaw, Poland} \\[-2mm]  }

\date{}

\begin{document} 

\maketitle 

\vfill

\begin{abstract}

Measurement of the 
\( \Gamma (h\rightarrow \gamma \gamma ){\rm {Br}}(h\rightarrow b\overline{b}) \) 
at the Photon Collider at TESLA is studied 
for the Standard Model Higgs boson with mass of 120 to 160 GeV.
The NLO estimation of background, analysis of overlaying events, 
realistic $b$-tagging and corrections for escaping neutrinos were performed.
We find that for \( M_{h}= \) 120-160 GeV
the \( \Gamma (h\rightarrow \gamma \gamma ){\rm Br}(h\rightarrow b\overline{b}) \)
can be measured with a statistical accuracy of 2-7\% 
after one year of the Photon Collider running.

\end{abstract}


\section{Introduction}

A  search of the Higgs boson is among the most important tasks for the present and
future colliders. Once the Higgs boson is discovered, it will be crucial
to determine its properties with a high accuracy.  
A photon-collider option of the TESLA
 collider \cite{TDR} offers a  unique possibility to produce
the Higgs boson as an \( s \)-channel resonance. The neutral Higgs boson
couples to the photons  through a loop 
with the massive charged particles. This loop-induced $ h \gaga $ coupling
is sensitive to contributions of new particles which appear in various 
extensions of  the SM. 

The SM Higgs boson with a  mass below \( \sim 140\) GeV is expected
to decay predominantly into the \( \bbbar \) final state. 
Here we consider the process \( \gagahbb \) 
for a Higgs-boson mass  \Mheq 120, 130, 140, 150 and 160 GeV
at the Photon Collider at TESLA.
Both the signal and  background
events are generated according to a  realistic photon--photon luminosity
spectrum \cite{V.Telnov}, parametrized by a CompAZ model \cite{CompAZ}. 
For the first time in such study 
overlaying events \gagahad{} are taken into account, 
for which we use photon--photon luminosity spectra from a full 
simulation \cite{V.Telnov}.
Our  analysis incorporates a  simulation
of the detector response according to the program SIMDET \cite{SIMDET401} 
and -- the next new element -- a realistic \btagging{} \cite{Btagging}.
This analysis supersedes our earlier analyses presented in \cite{NZKhbbm120appb,NZKpragacern,NZKamsterdam}.

\section{Photon--photon luminosity spectra}

The Compton back-scattering of a laser light off  high-energy
electron beams is considered as a  source of high energy, highly polarized
photon beams  \cite{Ilya}. 
According to the  current design \cite{TDR}, 
the energy
of the laser photons is assumed to be fixed
for all considered electron-beam energies;
laser photons are assumed to have circular polarization $P_{c} = $  100\%,
electrons longitudinal polarization is  $P_{e} = $ 85\%. 
We use the luminosity spectrum peaked at high energy and assume that 
the energy of primary electrons is adjusted in order to  enhance the signal
at a particular mass.

In a generation of the processes \gagahad{} one has to take into account 
also the low energy events, since they contribute to overlaying events \cite{TDR}.
 To simulate  them   we use  the  realistic $\gaga$ luminosity-spectra
for the photon collider at TESLA \cite{V.Telnov}, 
with the non-linear corrections and higher order QED processes.
For generation of the processes $\gagahbb$ and $\gagabbgccg$ we use the 
CompAZ parametrization \cite{CompAZ} of the spectrum \cite{V.Telnov}.

The results presented in this paper were obtained 
for an integrated luminosity
corresponding to one year of the Photon Collider running, as given by \cite{V.Telnov}.
For example, for \sqrtseeeq 210.5 GeV, 
which is optimal  for \Mheq 120 GeV,
the total photon--photon luminosity per year is $L_{\gaga}=410$ fb$^{-1}$
(84 fb$^{-1}$ for $ \Wgaga >80$ GeV). 
%
%
The total photon-photon luminosity increases  
to about 490 fb$^{-1}$ for \sqrtseeeq 260 GeV (used for \Mheq 160 GeV).

\section{Details of a simulation \\ and the first results for \Mheq 120 GeV }

We calculated the total width and  branching ratios of the SM Higgs
boson, using the program HDECAY \cite{HDECAY}, where
higher order QCD corrections are included. 
A generation of events was done with
the PYTHIA \pyver{} program \cite{PYTHIA}.
A parton shower algorithm, implemented in PYTHIA,
was used to generate the final-state particles. 

The  background events due to processes 
$\gagabbgccg$
were  generated using the program written by G.~Jikia \cite{JikiaAndSoldner},
where a complete  NLO QCD  calculation for the production of  massive
quarks is performed within the massive-quark scheme. 
The program includes exact one-loop QCD corrections to the lowest order
(LO) process
$\gagabbcc$
\cite{JikiaAndTkabladze}, and in addition 
the non-Sudakov form factor in the double-logarithmic
approximation, calculated up to four loops \cite{MellesStirlingKhoze}.

For an estimation of systematic uncertainties in  \btagging{} 
simulation we also generated the background events,
using the LO QED cross section for the processes
\( \gagabb \) and \( \gagacc \), and 
including parton shower, as  implemented in  PYTHIA.

The fragmentation into hadrons for all processes was performed using the PYTHIA program. 

%
%
\begin{figure}[t]
{\centering \resizebox*{!}{\figheight}%
            {\includegraphics{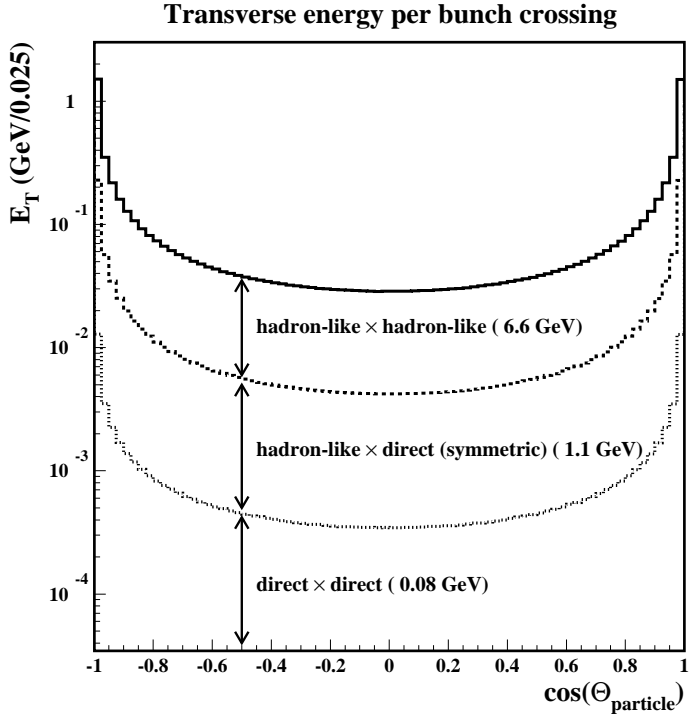}} \par}

\caption{\label{fig:GaGaHadAngET}
Angular distributions of transverse energy,  \( E_T \), for  \gagahad{} events per bunch crossing.
Various components and their sums are indicated. Generation was done for \sqrtseeeq 210.5 GeV~\cite{NZKamsterdam}. 
}
\end{figure}

%
Because of the large cross section, about one \gagahad{} event\footnote{%
We consider only photon--photon events with $\Wgaga >4 $ GeV.}  
is expected per bunch crossing at TESLA Photon Collider 
(for \sqrtseeeq 210-260 GeV, with nominal luminosity). 
We generate these events according to PYTHIA \pyver,
including direct and hadron-like photon contributions.
%
%
%
We use the full simulation of photon--photon 
spectra \cite{V.Telnov}, 
rescaled to the chosen beam energy.
For each considered collision energy, $\sqrtsee$, average number 
of \gagahad{} events per bunch crossing is calculated.  
Next, for each signal $\gagahbb$ or background $\gagabbgccg$ event 
\gagahad{} events are overlaid (added to the event record)
according to the Poisson distribution.  

Processes contributing to the overlaying events 
have forward-peaked distributions, as is shown in Fig.\ \ref{fig:GaGaHadAngET} for \sqrtseeeq 210.5 GeV. 
Therefore, to minimize an influence of these events on the signal measurement
we ignore
tracks and clusters with 
$|\cos(\theta_{i})|>\cos(\theta_{min})=0.9$ 
($\theta_{min}=450$ mrad; where the angle between the beam axis and a track/cluster, $\theta_{i}$,  
is measured in the laboratory frame). 
Hadron-level studies performed with PYTHIA show that after this cut
overlaying events contribute to the measured invariant mass 
below 5\% in 90\% of signal events.
Below we use the  $\theta_{min}$ cut  only when overlaying events are included 
in the analysis.
%
%

%
The fast simulation  program  SIMDET version 4.01 \cite{SIMDET401}
was used  to model a TESLA detector performance. 

Jets were reconstructed
using  the Durham algorithm, with \( y_{cut} = 0.02 \); the distance measure
was defined as 
\( y_{ij}=2\min (E^{2}_{i},E^{2}_{j})(1-\cos \theta _{ij})/E^{2}_{vis} \),
where $E_{vis}$ is the total energy measured in the detector.

The following cuts were used  to 
select the $\hbb$ events:
\begin{enumerate}

\item since the Higgs bosons are expected to be produced
  almost at rest, we require that the ratio of the total 
 longitudinal momentum of all observed particles 
 to the total visible energy is
\( |P_{z}|/E_{vis}<0.15 \),
\item we select two- and three-jet events, \( N_{jets}=2,\, 3 \), so that 
 events with one additional jet due to a hard-gluon emission are also accepted,
\item for each jet we require 
   \( |\cos \theta _{i}|<0.75 \), $i=1, ..., N_{jets}$.
\end{enumerate}

We use  ``ZVTOP-B-Hadron-Tagger'' package for 
the TESLA collider \cite{Btagging} for realistic \btagging{} simulation.
The package is based on the neural-network algorithm trained on the $Z$ decays. 
For each jet it returns a ``$b$-tag'' value -- the number 
between 0 and 1 corresponding to ``$b$-jet'' likelihood.
At \sqrtseeeq 210.5 GeV, the signal to the background ratio, 
$N( \gagabbg )/N( \gagaccg )$,
for selected 2-jet  events, after additional cut $E_{vis}>85$ GeV,
was investigated as a function of two $b$-tag values, 
with and without overlaying events.
%

\begin{figure}[htb]
{\centering \resizebox*{!}{\figheight}%
            {\includegraphics{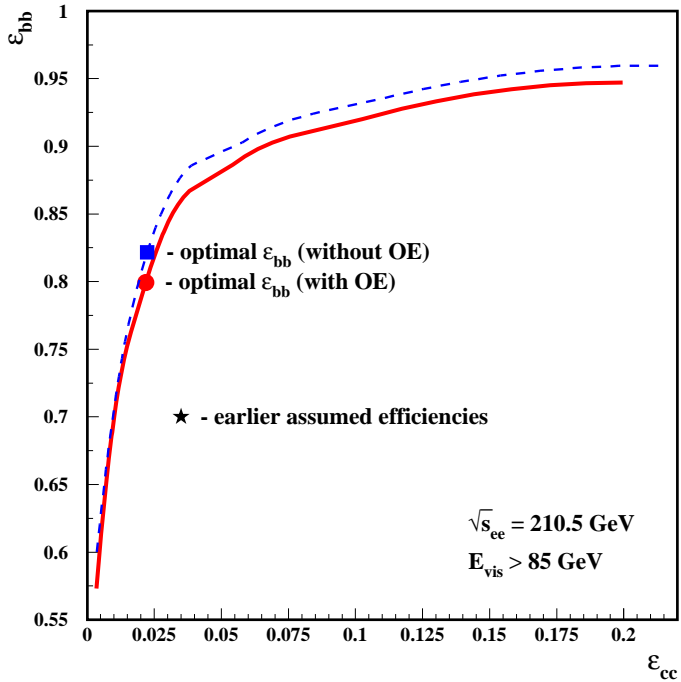}} \par}

\caption{\label{fig:BBtaggingCCtagging}
The \bbtagging{} efficiency of $ \gagabbg $ events, $\varepsilon_{bb}$,
versus
\ccmistagging{} probability of $ \gagaccg $ events, $\varepsilon_{cc}$, 
for \sqrtseeeq 210.5 GeV with the additional cut $E_{vis}>85$ GeV,
without and with  overlaying events. 
Optimal $\varepsilon_{bb}$ (and $\varepsilon_{cc}$) from these simulations (square and dot) 
and the earlier estimate (star) are indicated.
}
\end{figure}

By accepting $\gagabbgccg$ events above a given signal-to-background ratio,
the \bbtagging{} efficiency $\varepsilon_{bb}$ 
was studied as a function of  \ccmistagging{} probability $\varepsilon_{cc}$,
as shown in Fig.\ \ref{fig:BBtaggingCCtagging}.
For 3-jet events three possible pairs of jets were considered and 
the event was accepted if at least one pair gives 
the signal-to-background ratio above the cut.
%
It was found that  the cut corresponding to the efficiencies
$\varepsilon_{bb}=80\%$ and  $\varepsilon_{cc}=2.2\%$
is optimal for the $\Ghgagahbb$ measurement. 
For other considered electron-beam energies similar efficiencies were obtained (not shown).
In the earlier analyses \cite{JikiaAndSoldner,NZKhbbm120appb}
a fixed efficiency for the \bbtagging, 
  \( \varepsilon _{bb}=70\% \), and a fixed 
 probability  for a mistagging of the 
\( \ccbar \) events, \( \varepsilon _{cc}=3.5\% \), 
were assumed  (indicated in Fig.\ \ref{fig:BBtaggingCCtagging}).

%
In order to estimate possible influence of neglecting soft-gluon emissions
in simulation of a NLO background, for which parton shower algorithm is not used, 
the efficiency of \bbtagging{} was compared with an efficiency for 
the LO background simulation with PYTHIA (including parton shower) \cite{NZKpragacern}.
The overlaying events   were not included in this investigation.
Results of this analysis do not indicate any significant influence
on the \btagging{}. 
We obtain efficiencies of 
$\varepsilon_{bb}=82.2\% \, (82.4\%)$ and
$\varepsilon_{cc}=2.2\% \, (2.3\%)$, 
for background events generated according to NLO cross-sections without
parton shower (LO cross-sections with parton shower).
%

\begin{figure}[htb]
{\centering \resizebox*{!}{\figheight}%
            {\includegraphics{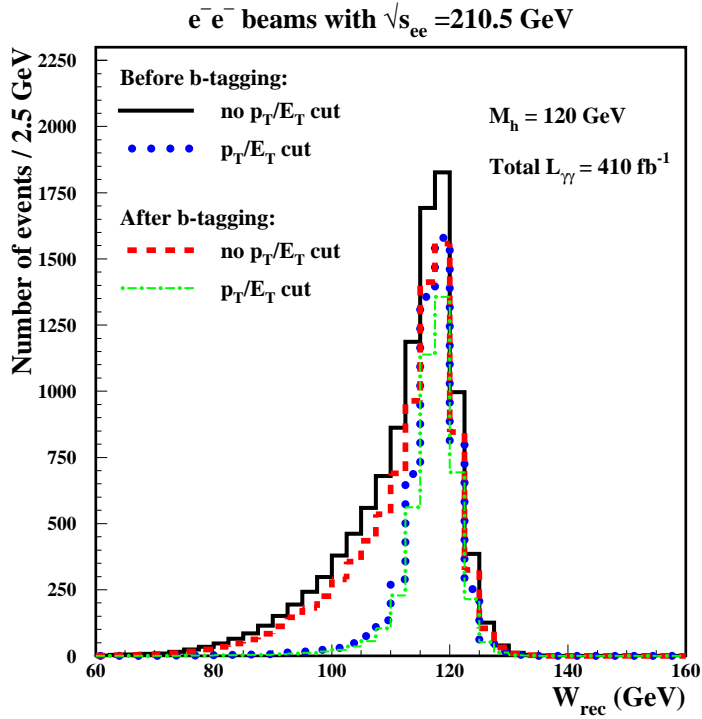}} \par}

\caption{\label{fig:HNoBtagBtag}
Reconstructed invariant mass, \protect\( W_{rec}\protect \),
distributions for selected $\gagahbb$ events, for \Mheq 120 GeV.
%
Distributions obtained before and after \btagging, without and with an additional \protect\( P_{T}/E_{T} < 0.04 \protect\)
cut are compared. 
%
}
\end{figure}

\begin{figure}[htb]
{\centering \resizebox*{!}{\figheight}%
            {\includegraphics{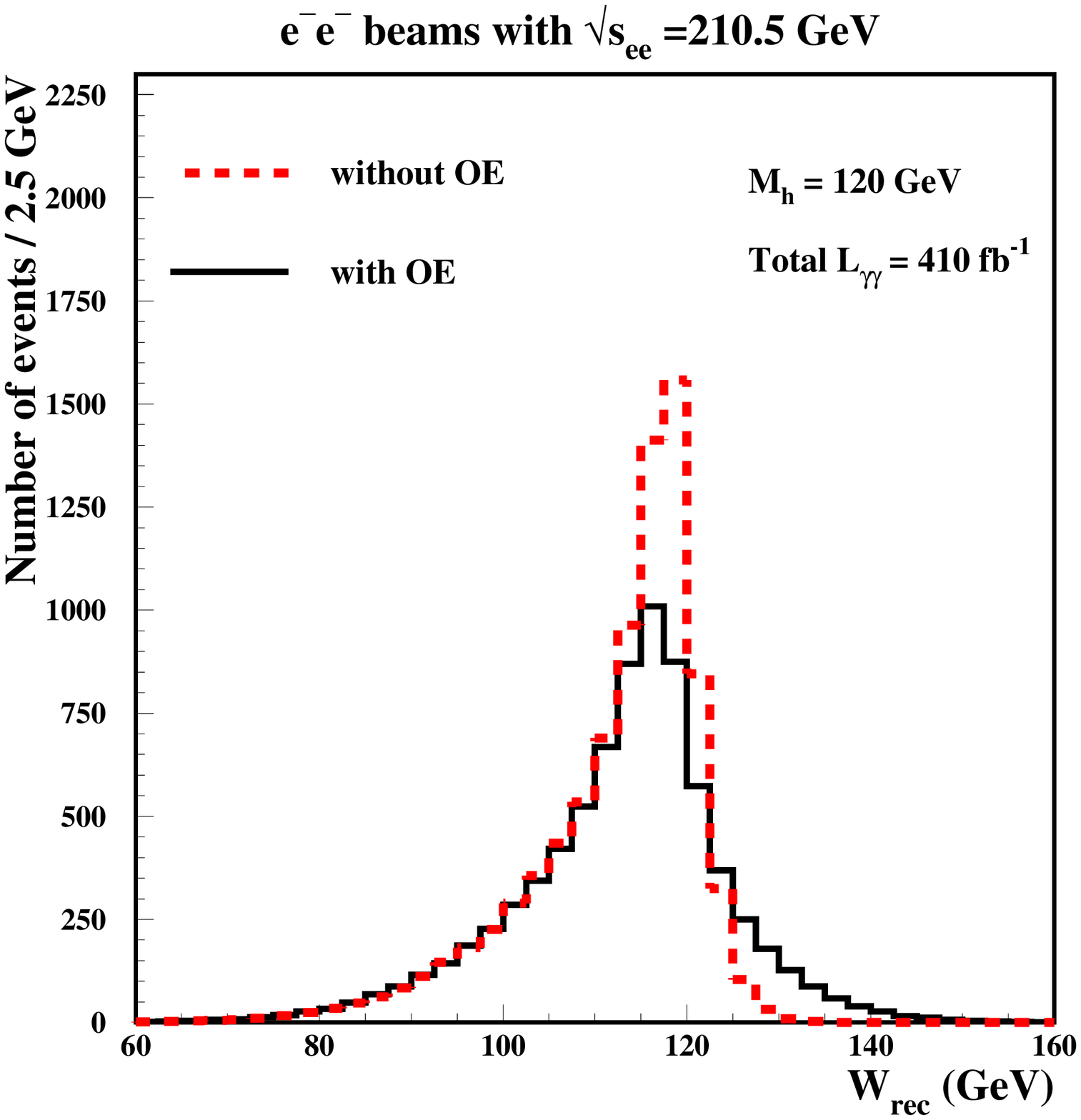}} \par}

\caption{\label{fig:HHOEBtag}
Reconstructed invariant-mass, \protect\( W_{rec}\protect \),
distributions for selected $\gagahbb$ events after \btagging, for \Mheq 120 GeV,
obtained 
without and with overlaying events (OE).
%
}
\end{figure}

In Fig.\ \ref{fig:HNoBtagBtag} we show influence of a \btagging{} on 
reconstructed invariant-mass distribution, \( W_{rec} \),
for the signal events $\gagahbb$. 
These results were obtained without overlaying events \gagahad{}. 
The low mass tail is due to the  presence of events with escaping neutrinos 
(see \cite{NZKhbbm120appb} for more details). 
Contribution of these events can be suppressed by an additional cut 
 \( P_{T}/E_{T} < 0.04 \), where
\( P_{T} \) and \( E_{T} \) are the absolute values of the total transverse
momentum of an event, $\vec{P}_{T}$, and the total transverse energy, 
respectively. We see that \btagging{} does not influence significantly
the shape of the distributions.
In Fig.\ \ref{fig:HHOEBtag} we compare the invariant-mass distributions 
before and after taking into account the overlaying events.
A mass resolution, derived from the 
Gaussian fit in the region from \( \mu - \sigma  \)
to \( \mu + 2 \sigma  \), is 3.5 and 6.1 GeV, respectively. 
Despite a quite high $\theta_{min}$ cut, the overlaying events result in  
a bigger tail above  \( W_{rec} = 120 \) GeV.
%
%
%
%
%
%
%
A small drop in a selection efficiency, 
resulting in the reduced number of events (from about 8520 to 7740 events), 
is observed. 
This is because the energy deposits from the \gagahad{} processes,
remaining after the $\theta_{min}$ cut, ``shift'' 
jets nearer to the beam axis and the event 
can be rejected by the jet-angle cut.
Moreover, the additional deposits and  $\theta_{min}$-cut deform jets, what 
slightly reduces the selection efficiency.
To study this issue in details we plan to simulate in the future 
the signal events with various $\theta_{min}$ values. 
%
%

\begin{figure}[htb]
{\centering \resizebox*{!}{\figheight}%
            {\includegraphics{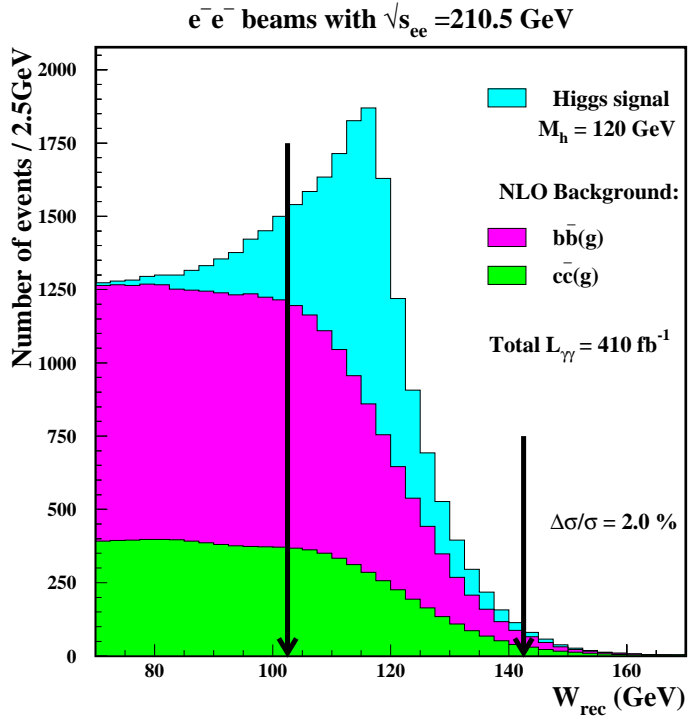}} \par}

\caption{\label{fig:ResultWithNLOBackgd}
Reconstructed invariant-mass, \protect\( W_{rec}\protect \),
distributions for selected $\bbbar$ events.
Contributions of the signal, for \Mheq  120 GeV, 
and of the heavy-quark background, calculated in the NLO QCD, are shown separately.
Arrows indicate the mass window, 102.5 to 142.5 GeV, optimized for the measurement of the 
$\Ghgagahbb$, which leads to the statistical precision of 2.0\%.
}
\end{figure}

After applying the selection cuts, \btagging{} and rejecting low-angle deposits we obtain the
distributions of the reconstructed \( \gaga  \) invariant
mass, \( W_{rec} \), 
shown in Fig.\ \ref{fig:ResultWithNLOBackgd}. 
The signal and NLO background contributions,
$\bbbar(g)$ and  $\ccbar(g)$, are shown separately.

Assuming that the signal for  Higgs-boson production  
will be extracted
by counting the number of \( \bbbar \) events in the mass window 
around the peak, $N_{obs}$, 
and subtracting the expected background events, $N_{bkgd}$,
we can calculate 
the expected relative statistical error for the cross section 
\( \sgagahbb  \)
(or the partial width multiplied by the branching ratio \( \Ghgagahbb  \))
in the following way:
\[
\frac{\Delta \sgagahbb}{\sgagahbb} =
\frac{\Delta \left[ \Ghgagahbb \right] }{\left[ \Ghgagahbb \right] }=
\frac{\sqrt{N_{obs}}}{N_{obs}-N_{bkgd}}.
\]
The accuracy expected for the  Higgs-boson mass of 120 GeV,
from  the reconstructed invariant-mass distribution
in the selected mass range between 102.5 and 142.5 GeV 
(see  Fig.\ \ref{fig:ResultWithNLOBackgd}),
is equal to 2.0\%. 
It is   in agreement with 
the results of a previous analysis \cite{NZKhbbm120appb}.
Note however, that in the present analysis a loss of signal efficiency 
due to the overlaying events 
is compensated by a more effective \btagging.

\section{Final results for masses 120--160 GeV}

\begin{figure}[htb]
{\centering \resizebox*{!}{\figheight}%
            {\includegraphics{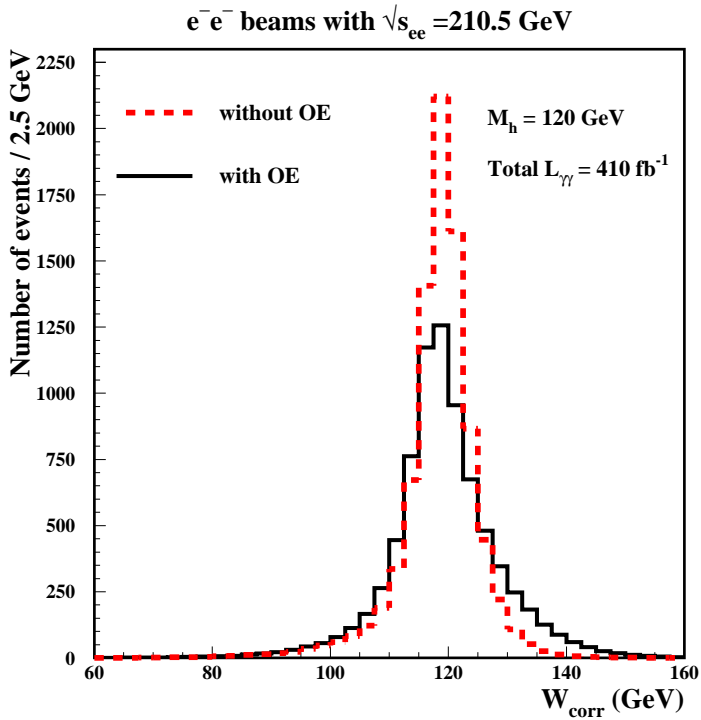}} \par}

\caption{\label{fig:HHOEBtagWcorr}
Corrected invariant mass, \protect\( W_{corr}\protect \),
distributions for selected $\gagahbb$ events after \btagging, for \Mheq 120 GeV,
obtained 
without and with overlaying events (OE).
%
}
\end{figure}

As in \cite{NZKhbbm120appb}, to correct for escaping neutrinos
we use 
the corrected invariant mass, a variable defined as: 
\begin{equation}
W_{corr} \equiv \sqrt{W^{2}_{rec}+2P_{T}(E_{vis}+P_{T})}
\end{equation}
In Fig.\ \ref{fig:HHOEBtagWcorr} the distributions of \( W_{corr} \)
for the selected signal events, without and with overlaying events, 
are presented. 
The tail of events with invariant masses below $\sim 110$ GeV 
is much smaller than for the $W_{rec}$-distributions (compare with Fig.\ \ref{fig:HHOEBtag}). 
The mass resolutions, 
derived from the Gaussian fits to the $W_{corr}$-distributions 
in the region from \( \mu - 2 \sigma  \) to \( \mu + \sigma  \),
are equal to 3.5 and 5.0 GeV, 
without and with overlaying events,  respectively. 
%

\begin{figure}[htb]
{\centering \resizebox*{!}{\figheight}%
               {\includegraphics{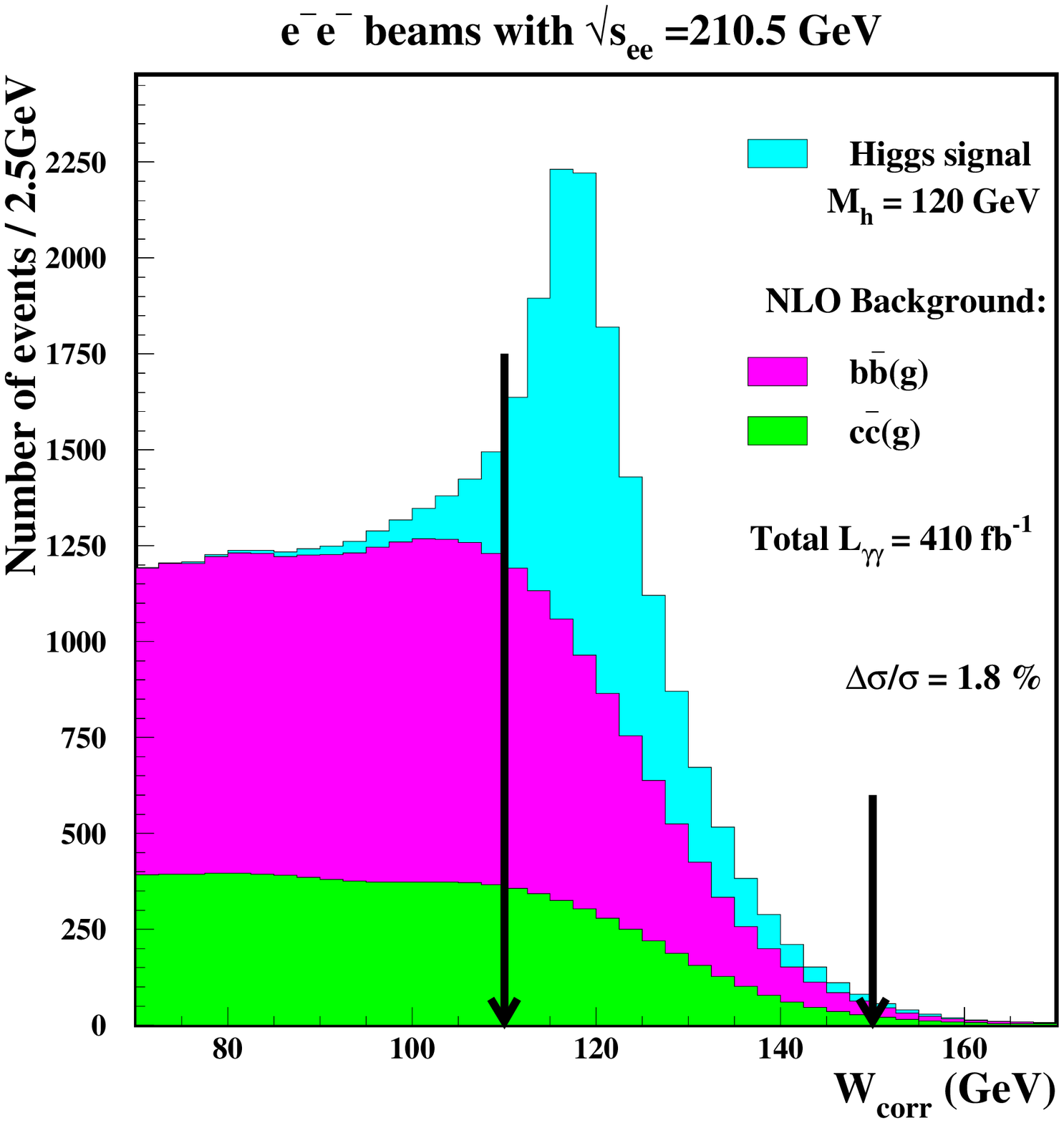}} \par}

\caption{\label{fig:Wcorr}
As in Fig.\ \ref{fig:ResultWithNLOBackgd}, for the corrected invariant mass, \protect\( W_{corr}\protect \),
distributions.
}
\end{figure}

The distributions of the \( W_{corr} \), obtained for the  signal and
background events, with overlaying events included,  are shown in Fig.\ \ref{fig:Wcorr}. 
The most precise measurement of the Higgs-boson cross section
is obtained for the  mass window \( W_{corr} \)
between 110 and 150 GeV, as indicated by arrows.
In the selected \( W_{corr} \) region one expects, after one year of
the Photon Collider running at nominal luminosity,
about 6900 reconstructed signal
events and 8800 background events  (\ie \( S/B \approx 0.8 \)).
This corresponds to the statistical precision of:
\[
\frac{\Delta \left[ \Ghgagahbb \right] }{\left[ \Ghgagahbb \right] }=1.8\%. 
\]
%

\begin{figure}[thb]
{\centering \resizebox*{!}{\figheight}%
               {\includegraphics{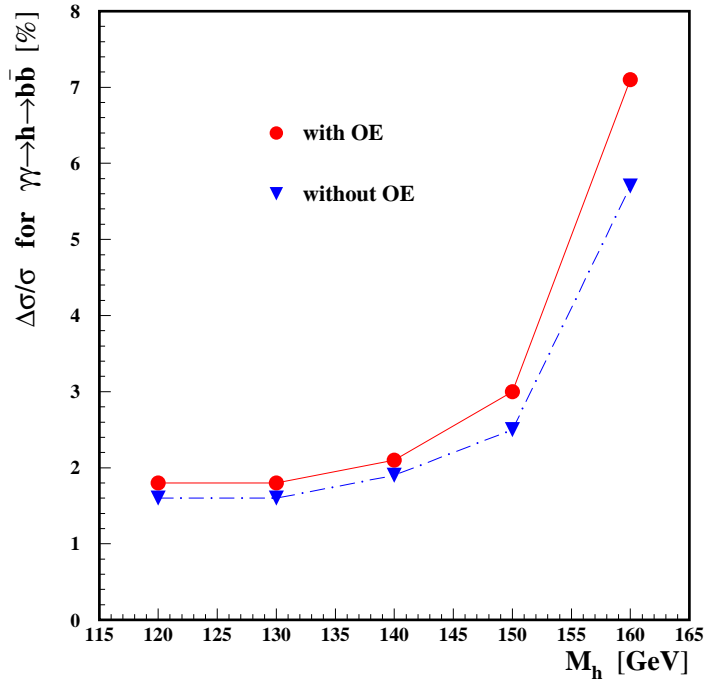}} \par}

\caption{\label{fig:PrecisionSummary}
Statistical precisions of $\Ghgagahbb$ measurements for the SM Higgs boson with mass 120-160 GeV
with and without overlaying events (OE).
The lines are drawn to guide the eye. 
}
\end{figure}

%
We have performed a full simulation of signal and background events also
for \Mheq  130, 140, 150 and 160 GeV
choosing optimal $\emem$ beam energies for each Higgs-boson mass.
Statistical precision of $\Ghgagahbb$ measurement was estimated in each case.
It is equal to 1.8\%, 2.1\%, 3.0\% and 7.1\%, respectively.
These results, together with described above result for \Mheq 120 GeV, are presented in Fig.\ \ref{fig:PrecisionSummary}. 
For comparison our earlier results \cite{NZKamsterdam}, 
obtained without overlaying events, are also shown.

\section{Conclusions}

We performed a realistic simulation of SM Higgs-boson production 
in the Photon Collider at TESLA, $\gagahbb$,
with the NLO background, corrections for escaping neutrinos
and --  for the first time -- with the realistic \btagging{} and overlaying events.
Our analysis shows that for \Mheq 120-160 GeV
the two-photon width of SM Higgs boson can be measured
with a statistical precision  1.8-7.1\%.
The obtained accuracies
are in a rough agreement with the results of a previous analysis, based on the
idealistic  Compton spectrum \cite{JikiaAndSoldner}.
As shown in  \cite{NZKhbbm120appb}, the realistic photon--photon luminosity spectrum 
is more challenging for a precise determination of $\Ghgagahbb$.
The precision of the measurement has been improved
after applying a correction for escaping neutrinos  
and a mass-window cut. 
In the present analysis we include overlaying events
and, as expected, the measurement precision decreases.
However, a realistic \btagging{} used here,
resulting in a  better \bbtagging{} efficiency than  estimated earlier,
allows to counterbalance this effect.

%
The measurement discussed in this paper can be used to derive the partial width
\( \Gamma (h \rightarrow \gaga ) \) \cite{TDR}.
%
%
For example, for a Higgs boson of \Mheq 120  GeV we estimate a precision 1.8\% 
on the measurement of \( \Ghgagahbb \).
Assuming that \( {\rm Br}(h \rightarrow \bbbar) \)  will be measured 
at the $\epem$ Linear Collider with a precision 1.5\% \cite{Brient}, 
the partial width \( \Gamma (h\rightarrow \gaga ) \) can be extracted with an accuracy of 2.3\%.
Using in addition the result from the $\epem$ Linear Collider for 
\( {\rm Br}(h \rightarrow \gaga) \) 
\cite{Boos}, one can also extract 
\( \Gamma_{\rm tot} \) with a precision of 10\%.

For higher masses of the SM Higgs boson other decay channels
should be considered, see \eg \cite{wwzz}.

\subsection*{Acknowledgments}

We would like to thank  A.\ De Roeck, K.\ M\"onig and T.~Kuhl for valuable discussions.
M.K.~acknowledges partial
support by the Polish Committee for Scientific Research, Grants 2 P03B 05119 (2003), 
5 P03B 12120 (2003), and by the European Community's
Human Potential Programme under contract HPRN-CT-2000-00149 Physics
at Colliders.


\begin{thebibliography}{1}

\bibitem{TDR}B.~Badelek et al., TESLA Technical Design Report, Part VI, Chapter
1: The Photon Collider at TESLA, DESY 2001-011, ECFA 2001-209, hep-ex/0108012.

\bibitem{V.Telnov}V.~Telnov, Nucl.~Instrum.~Meth.~A 355, 3 (1995); 
V.~Telnov, \textit{A code PHOCOL for the simulation of luminosities 
and backgrounds at photon colliders}, 
talk presented at the Second Workshop of ECFA-DESY Study,
Saint~Malo, France, April 2002, 
http://www.desy.de/\textasciitilde{}telnov/stmalo/stmalo1.ps.gz;
for details of the simulation see also: 
http://www.desy.de/\textasciitilde{}telnov/ggtesla/spectra/.

\bibitem{CompAZ}A.F. \.Zarnecki, \textit{CompAZ: parametrization of the 
photon collider~luminosity~spectra},\\ 
Acta Phys.~Polon.~B34, 2741-2758 (2003), hep-ex/0207021; \\ 
http://info.fuw.edu.pl/\textasciitilde{}zarnecki/compaz/compaz.html.




\bibitem{SIMDET401}M.~Pohl and H.~J.~Schreiber,
DESY-02-061, hep-ex/0206009.

\bibitem{Btagging}T.~Kuhl and K.~Harder, 
\textit{B-tagging in SIMDET: 1st application to Higgs}, 
talk presented at the Second Workshop of ECFA-DESY Study,
Saint~Malo, France, April 2002, 
%
http://www-dapnia.cea.fr/ecfadesy-stmalo/Sessions/Higgs/session1/kuhl.ps.


\bibitem{NZKhbbm120appb}
P. Nie\.zurawski, A.F. \.Zarnecki, M. Krawczyk,  
\textit{The SM Higgs-boson production 
in \( \gamma \gamma \rightarrow h\rightarrow b\bar{b} \)
at the Photon Collider at TESLA},
Acta Physica Polonica B 34 177-187 (2003),
CERN-TH-2002-166, IFT-2002-29, \mbox{hep-ph/0208234}. 

\bibitem{NZKpragacern}
P. Nie\.zurawski, A.F. \.Zarnecki, M. Krawczyk,  
\textit{B-tagging and the light Higgs-boson production at Photon Collider at TESLA}, 
talk presented at the 3rd Workshop of the Extended
ECFA/DESY Study, Prague, Czech Repulic, November 2002;
P. Nie\.zurawski, A.F. \.Zarnecki, M. Krawczyk,  
\textit{B-tagging uncertainty due to Parton Shower}, 
talk presented at the Gamma--Gamma Physics Meeting, CERN, February 2003.

\bibitem{NZKamsterdam}
P. Nie\.zurawski, A.F. \.Zarnecki, M. Krawczyk,  
\textit{New results for $ \gamma  \gamma \rightarrow H \rightarrow b \bar{b} $ in SM and MSSM}, 
talk presented at the 4th ECFA/DESY Workshop, Amsterdam, Netherlands,  April 2003.




\bibitem{Ilya}
I.~F. Ginzburg, G.~L. Kotkin, V.~G. Serbo and V.~I. Telnov,
\newblock { Pizma~ZhETF\/} 34, 514 (1981),
\newblock {JETP Lett.} 34, 491 (1982);
 Preprint INP 81-50, Novosibirsk, 1981 and 
\newblock Nucl.~Instrum.~Meth.~A 205, 47 (1983);
\newblock Preprint INP 81-102, Novosibirsk, 1981;
\\
I.~F. Ginzburg, G.~L. Kotkin, S.~L. Panfil, V.~G. Serbo and V.~I. Telnov,
\newblock { Nucl.~Instrum.~Meth.\/} A219, 5 (1984);
\\
V.~I. Telnov,
\newblock { Nucl.~Instrum.~Meth.\/} A294, 72 (1990) and  A355, 3 (1995).

\bibitem{HDECAY}A.~Djouadi, J.~Kalinowski and M.~Spira, Comput.~Phys.~Commun.~108, 56
(1998), \mbox{hep-ph/9704448}.

\bibitem{PYTHIA}T.~Sj\"ostrand, P.~Eden, C.~Friberg, L.~Lonnblad, G.~Miu, S.~Mrenna
and E.~Norrbin, Comput.~Phys.~Commun.~135, 238 (2001), hep-ph/0108264.

\bibitem{JikiaAndSoldner}G.~Jikia and S.~S\"oldner-Rembold, Nucl.~Instrum.~Meth.~A 472, 133 (2001), hep-ex/0101056. S.~S\"oldner-Rembold, talk presented at the 10th International Conference on Supersymmetry and Unification of Fundamental Interactions (SUSY02), Hamburg, Germany, Jun 2002. 

\bibitem{JikiaAndTkabladze}G.~Jikia and A.~Tkabladze, Nucl.~Instrum.~Meth.~A 355, 81 (1995) and  Phys.~Rev.~D 54, 2030 (1996), hep-ph/9406428.

\bibitem{MellesStirlingKhoze}M.~Melles and W.~J.~Stirling, 
Phys.~Rev.~D 59, 94009 (1999) and  Eur.~Phys.~J.~C
9, 101 (1999), hep-ph/9807332; M.~Melles, W.~J.~Stirling and V.~A.~Khoze,
Phys.~Rev.~D 61, 54015 (2000), hep-ph/9907238; M.~Melles, Nucl.~Instrum.~Meth.~
A 472, 128 (2001), hep-ph/0008125.



\bibitem{Brient}
J-C.~Brient,
\newblock  LC-PHSM-2002-003; 
M.~Battaglia, 
\newblock  hep-ph/9910271.



\bibitem{Boos}
E.~Boos, J.~C.~Brient, D.~W.~Reid, H.~J.~Schreiber and R.~Shanidze,
Eur.\ Phys.\ J.\ C {\bf 19} 455 (2001), hep-ph/0011366.


\bibitem{wwzz}
P. Nie\.zurawski, A.F. \.Zarnecki, M. Krawczyk,  
\textit{Study of the Higgs-boson  \mbox{decays} into 
       $W^+W^-$ and $Z Z $ at the Photon Collider},
   J.~High Energy Phys. 11 034 (2002), 
   CERN-TH/2002-165, IFT-28/2002, \mbox{hep-ph/0207294}.

%
\end{thebibliography}
\end{document}